\documentclass[aoas,MSNbibl,nameyear,dvips]{arximspdf}
\usepackage{dcolumn}
\usepackage{graphicx}

\doi{10.1214/14-AOAS756} 
\volume{8}
\issue{3}
\pubyear{2014}
\firstpage{1782}
\lastpage{1799}
\docsubty{FLA}

\makeatletter
\newcolumntype{d}[1]{D{.}{.}{#1}}
\makeatother

\begin{document}
\begin{frontmatter}

\title{Incorporating geostrophic wind information for improved
space--time short-term wind speed~forecasting\thanksref{T1}}
\runtitle{Improved wind speed forecasting}

\begin{aug}
\author[A]{\fnms{Xinxin}~\snm{Zhu}\ead[label=e1]{email2xzhu@gmail.com}\thanksref{T2,m1}},
\author[B]{\fnms{Kenneth P.}~\snm{Bowman}\ead[label=e2]{k-bowman@tamu.edu}\thanksref{m1}}
\and
\author[C]{\fnms{Marc G.}~\snm{Genton}\corref{}\ead[label=e3]{marc.genton@kaust.edu.sa}\thanksref{T2,m2}}
\runauthor{X. Zhu, K. P. Bowman and M. G. Genton}
\affiliation{Texas A\&M University\thanksmark{m1} and King Abdullah University of Science~and~Technology\thanksmark{m2}}
\address[A]{X. Zhu\\
Department of Statistics\\
Texas A\&M University\\
College Station, Texas 77843-3143\\
USA\\
\printead{e1}} 
\address[B]{K. P. Bowman\\
Department of Atmospheric Sciences\\
Texas A\&M University\\
College Station, Texas 77843-3150\\
USA\\
\printead{e2}}
\address[C]{M. G. Genton\\
CEMSE Division\\
King Abdullah University of Science\\
\quad and Technology\\
Thuwal 23955-6900\\
Saudi Arabia\\
\printead{e3}}
\end{aug}
\thankstext{T1}{This publication is partly based
on work supported by Award No. KUS-C1-016-04 made by King Abdullah
University of Science and Technology (KAUST).}
\thankstext{T2}{Supported in part by NSF Grant DMS-10-07504.}

\received{\smonth{6} \syear{2013}}
\revised{\smonth{5} \syear{2014}}

%
\begin{abstract}
Accurate short-term wind speed forecasting is needed for the rapid
development and efficient operation of wind energy resources. This is,
however, a very challenging problem. Although on the large scale, the
wind speed is related to atmospheric pressure, temperature, and other
meteorological variables, no improvement in forecasting accuracy was
found by incorporating air pressure and temperature directly into an
advanced space--time statistical forecasting model, the trigonometric
direction diurnal (TDD) model. This paper proposes to
incorporate the geostrophic wind as a new predictor in the TDD model.
The geostrophic
wind captures the physical relationship between wind and pressure
through the observed approximate balance between the pressure gradient
force and the Coriolis acceleration due to the Earth's rotation.
Based on our numerical experiments with data from West Texas, our new
method produces more accurate forecasts than does the TDD model using
air pressure and temperature for 1- to 6-hour-ahead forecasts based on three
different evaluation criteria. Furthermore, forecasting errors can be
further reduced by using moving average hourly wind speeds to fit the
diurnal pattern. For example, our new method obtains between 13.9\%
and 22.4\% overall mean absolute error reduction relative to
persistence in 2-hour-ahead forecasts,
and between 5.3\% and 8.2\% reduction relative
to the best previous space--time methods in this setting.
\end{abstract}

%
\begin{keyword}
\kwd{Geostrophic wind}
\kwd{space--time statistical
model}
\kwd{wind energy}
\kwd{wind speed forecasting}
\end{keyword}
\end{frontmatter}

\section{Introduction}\label{sec1}

Because it is a rich resource that is both green and renewable, wind
energy has been developing rapidly worldwide; see the book
by \citet{haugen2012} on renewable energy and the reviews by \citet{genton2007},
\citet{zhu2012}, and \citet{pin2013} for more information about wind energy.

Wind power cannot be simply added into current power systems. Rather,
its introduction   creates costs and inefficiencies in power
systems. Because of the high uncertainties, nondispatchable and
limited predictability of wind energy, an increase in the proportion of
wind power in a system requires a corresponding increase of fast but
expensive nonwind backup power to balance wind fluctuations.

The solution to reducing the uncertainties of wind power generation is
accurate wind forecasting. In particular, short-term forecasting up to
a few hours ahead is essential. Long-term wind forecasting is less
accurate, while high-quality short-term prediction is possible.
In order to describe the uncertainty in the forecast of a future event,
the forecast ought to be probabilistic, that is, in the form of a predictive
probability distribution; see \citet{gneiting2014} for an overview.

At the same time, short-term forecasting is closely related to a power system
dispatch. In a power market, one-day-ahead, hours-ahead, and even
minutes-ahead price adjustments are used to determine how much
electricity each power plant should generate to meet demand at minimum
cost; see \citet{xie2011}. Moreover, if there is a gap between the
demand and the estimated supply, there is enough time to draw on less
expensive backup power plants. Accurate short-term forecasts reduce
the cost for reserves and stabilize the power system.

A number of short-term, statistical, wind forecasting models have been
developed; see reviews by \citet{giebel2011}, \citet{kar2004},
\citet{anl2009}, \citet{zhu2012} and \citet{pin2013}. Statistical
space--time forecasting
models that take into account both spatial and temporal correlations
in wind have been found to be particularly accurate for short-term
forecasting problems. The regime-switching space--time diurnal (RSTD)
models, proposed by \citet{gneiting2006}, were found to outperform
persistence (PSS), autoregressive and vector autoregressive models. Since
the RSTD models were introduced, researchers have sought to generalize
and improve them. For example, \citet{hering2010} proposed the
trigonometric direction diurnal (TDD) model to generalize the RSTD
model by treating wind direction as a circular variable and including
it in their model. \citet{zhu2013} generalized the RSTD model by
allowing forecasting regimes to vary with the prevailing wind and
season, obtaining comparable forecasting accuracy. They referred to their
model as a rotating RSTD model. \citet{pinson2012} used a first-order
Markov chain to determine the regime sequence in offshore wind power
forecasting problems and proposed the so-called adaptive
Markov-switching autoregressive models.

All of the aforementioned statistical wind forecasting models use only
historical wind information---wind speed and direction---to predict
future winds. Other atmospheric parameters, such as temperature and
pressure, are closely tied to the wind through various physical
processes and could potentially be included in models to improve
prediction accuracy. Directly incorporating temperature and pressure as
statistical predictors turns out not to be helpful, however, because
winds, for example, are related more closely to horizontal gradients of
pressure rather than pressure itself. Outside the tropics, the wind
field is closely tied to the large-scale atmospheric pressure field
through a balance between the horizontal pressure gradient force and
the Coriolis acceleration from the Earth's rotation. This relationship
is known as geostrophic balance [e.g., \citet{wallace2006}, Section~7.2]. Because of the physical relationship between pressure gradients
and winds, the atmospheric pressure field contains information about
the wind that is not contained in surface wind measurements. For a
general introduction   to meteorological basics of wind power generation,
see \citet{emeis2013} and the references therein. In this paper, a new
predictor is introduced to the TDD model, the geostrophic wind (GW),
which is the theoretical horizontal wind velocity that exactly balances
the observed pressure gradient force. This new model is named TDDGW.

Numerical experiments applying the TDDGW model to data from West Texas
are carried out for 1- to 6-hour-ahead wind
forecasting. The geostrophic wind direction (D) and the difference in
temperature (T) between the current and previous day are also
considered, with corresponding models named TDDGWD and TDDGWT,
respectively. Additionally, simpler but more efficient methods are
proposed to fit the prevailing diurnal wind pattern to obtain better
forecasts. Mean absolute errors (MAE), root mean squared errors (RMSE) and
continuous ranked probability scores (CRPS), as well as probability
integral transform histograms, are used to evaluate the
performance of the forecasting models.

The remainder of this paper is organized as follows. In Section~\ref{geowindsec.geowind} the geostrophic wind
estimation procedure is briefly introduced. In Section~\ref{geowindsec.tddgw} the TDDGW model is proposed, along with the
TDDGWD and TDDGWT models and modified diurnal pattern fitting methods.
The West Texas data are used as a case study
in Section~\ref{geowindsec.data}. In Section~\ref{geowindsec.results} forecast results are evaluated and compared
with those from reference models. Section~\ref{geowindsec.conclusion} offers final remarks. The
abbreviations used in the paper are listed in
Table~\ref{tab.abb}.
%
\begin{table}[t]
\textwidth=280pt
\caption{List of abbreviations}
\label{tab.abb}
\begin{tabular*}{\textwidth}{@{\extracolsep{\fill}}@{}lp{215pt}@{}}
\hline
$y$ & Wind speed\\
$\theta$ & Wind direction\\
$w_g$ & Geostrophic wind speed\\
$\theta_g$ & Geostrophic wind direction\\
PSS & Persistence \\
RSTD & Regime switching space--time diurnal model\\
TDD & Trigonometric direction diurnal model\\
TDDGW & TDD model incorporating geostrophic wind information\\
TDDGWT & Including 24-hour temperature difference into TDDGW\\
TDDGWD & Including geostrophic wind direction into TDDGW\\
TDDGWDT & Including 24-hour temperature difference and geostrophic wind direction into TDDGW\\
YMD & Modified diurnal pattern fitted with yearly period\\
SMD & Modified diurnal pattern fitted with seasonal period\\
MD & Modified diurnal pattern fitted with 45 days' period\\
\hline
\end{tabular*}
\end{table}

\section{Estimating the geostrophic wind}\label{geowindsec.geowind}

Using pressure as a vertical coordinate, the eastward and northward
components of the
geostrophic wind, $u_g$ and $v_g$, are given by
\begin{equation}
\label{geostrophic3} u_g = -\frac{g_0}{f} \frac{\partial Z}{\partial y} \quad
\mbox{and} \quad v_g = \frac{g_0}{f} \frac{\partial Z}{\partial x},
\end{equation}
where $x$ and $y$ are local eastward and northward Cartesian
coordinates, $g_0$ is the acceleration of gravity, $f=2 \Omega\sin
\phi
$ is the Coriolis parameter, $\phi$ is latitude, $\Omega$ is the
rotation rate of the Earth, and $Z$ is the height of a convenient
nearby surface of constant pressure.

For the region covered by this study, differences in $Z$ between
stations are small [$O$(10 m)] compared to the magnitude of $Z$
[$O$(1000 m)], so care must be taken to remove systematic biases and
noise in individual measurements of $Z$ to accurately estimate the
horizontal pressure gradient. To compute the geostrophic wind
components from a network of surface pressure observing stations, the
following steps are used. First, because the barometers at different
stations are typically located at different elevations above sea level,
it is necessary to adjust the pressure measurements to a standard
reference pressure. This can be done with good accuracy through the
hydrostatic equation, which in integral form is written as
%
\begin{equation}
\label{hypsometric} Z = Z_i + \frac{R\bar{T}}{g_0} \ln\biggl(
\frac{p_i}{p_{\mathrm{ref}}}\biggr),
\end{equation}
where $Z_i$ is the geopotential height of barometer $i$, $p_i$ is the
pressure measurement by barometer $i$, $p_{\mathrm{ref}}$ is the desired
reference pressure level (e.g., 850 hPa), $Z$ is the unknown
geopotential height of the reference pressure level, $R$ is the gas
constant for air (287 J K$^{-1}$ kg$^{-1}$), and $\bar{T}$ is the
layer-averaged temperature between $p_i$ and $p_{\mathrm{ref}}$, which in this
paper is estimated using surface temperature measurements.

Because $\Delta Z/Z \ll 1$ for horizontal scales of interest in this
study, systematic biases and random noise in the barometers would lead
to large errors in estimates of the pressure gradient. Biases are
removed by subtracting the time-mean pressure at each station for the
time series. This will also remove any real time-mean geostrophic wind,
but for statistical wind forecasting purposes, only variations in the
geostrophic wind are of interest. Random noise in the pressure
measurements are removed by fitting a smooth (planar) surface,
%
\begin{equation}
\label{plane} Z(x, y) = a_0 + a_1 x + a_2
y,
\end{equation}
to the geopotential heights at each time. From this, we get
\[
\frac{\partial Z}{\partial x} = a_1 \quad \mbox{and} \quad \frac{\partial Z}{\partial y} =
a_2,
\]
which can be substituted into equation (\ref{geostrophic3}) to give
%
\begin{equation}
\label{coefficients} u_g = -\frac{g_0}{f} a_2 \quad
\mbox{and}\quad v_g = \frac{g_0}{f}a_1.
\end{equation}
The geostrophic wind speed and direction are given by $w_g = \sqrt
{u_g^2 + v_g^2}$ and $\theta_g = \tan^{-1}(v_g/u_g)$,
respectively.

\section{The trigonometric direction diurnal model with geostrophic
wind}\label{geowindsec.tddgw}

\subsection{The TDD model and reference models}\label{refforc}

The TDD model [\citet{hering2010}] is an advanced space--time model for
short-term wind speed forecasting problems. It generalizes the RSTD
model [\citet{gneiting2006}] by treating wind direction as a circular
variable and including it in the model, such that the alterable and
locally dependent forecasting regimes are eliminated. The main idea of
this model is presented in this section in order to develop our new
model.

Let $y_{s,t}$ and $\theta_{s,t}$, $s = 1,\ldots,S$ and $t = 1,\ldots,T$,
be surface wind speed and direction measurements at station $s$ at
time $t$, respectively. The objective is to predict the $k$-step-ahead
wind speed, $y_{i,t+k}$, at one of the stations, $i \in\{1,\ldots, S\}$.
For short-term wind speed forecasting problems, the $k$-step-ahead
is from 1 to 6 hours.

Like \citet{gneiting2006} and \citet{hering2010},
it is assumed in the TDD model that $y_{s,t+k}$ follows a truncated
normal distribution, $N^+(\mu_{s,t+k},\allowbreak \sigma_{s,t+k})$, with
$\mu_{s,t+k}$ and $\sigma_{s,t+k}$ as the center parameter and the
scale parameter, respectively, considering that the density of the
wind speed is nonnegative. Of course, there are other alternative
probability distributions to fit wind speed, such as the Weibull,
Rayleigh and Beta distributions; see \citet{monahan2006}, \citet{monahan2011} and \citet{zhu2012}.

If the two parameters of the truncated normal distribution are modeled
appropriately, accurate probabilistic forecasts can be achieved beyond
point forecasts. In the TDD model, these two parameters are modeled as
follows, taking $s = 1$ as an example:

\begin{longlist}[(a)]
\item[(a)] The center parameter, $\mu_{1,t+k}$, is modeled in two parts:
\[
\mu_{1,t+k} = D_{1,t+k} + \mu^r_{1,t+k}.
\]
The first part, $D_{1,t+k}$, is the diurnal component in the wind speed,
which is fitted by two pairs of trigonometric functions:
%
\begin{eqnarray}
\label{geowindeq.D}
D_{1,h} &=&  d_0 + d_1 \sin \biggl(
\frac{2\pi h}{24} \biggr) +d_2\cos \biggl(\frac
{2\pi h}{ 24} \biggr)
\nonumber
\\[-8pt]
\\[-8pt]
\nonumber
&&{}+
d_3\sin \biggl(\frac{4\pi h}{24} \biggr) + d_4 \cos
\biggl(\frac{4\pi h}{ 24} \biggr),
\end{eqnarray}
where $h = 1, 2, \ldots, 24$.

The residual of the wind speed after removing the diurnal component is
modeled as
%
\begin{eqnarray}
\label{geowindeq.mur}
\mu^r_{1,t+k} &=& \alpha_0 + \sum_{s=1,\ldots,S} \biggl[ \sum_{j=0,1,\ldots,q_s}
\alpha_{s,j} y^r_{s,t-j}
\nonumber
\\[-8pt]
\\[-8pt]
\nonumber
&&\hphantom{\alpha_0 + \sum
_{s=1,\ldots,S} \bigg[}{}+ \sum
_{j'=0,1,\ldots,q_s'} \bigl\{ \beta _{s,j'}\cos\bigl(
\theta^r_{s,t-j'}\bigr) + \gamma_{s,j'}\sin\bigl(\theta
^r_{s,t-j'}\bigr) \bigr\} \biggr].\hspace{-25pt}
\end{eqnarray}

Equation (\ref{geowindeq.mur}) models the $k$-step-ahead wind speed residual
as a linear combination of current and past wind speed residuals at
all stations up to time lag $q_s$ depending on station $s$, as well as
a pair of trigonometric functions of wind direction residuals whose
diurnal patterns are also fitted by the model in (\ref{geowindeq.D})
up to
time lag $q'_s$, which is not necessarily equivalent to $q_s$. Both $q_s$
and $q'_s$ are determined by the modified Bayesian information criterion
(BIC) as described by \citet{hering2010}.

\item[(b)] The scale parameter is modeled by a simple linear model of
volatility value, $v^r_{t} $, in the following form:
\[
\sigma_{1,t+k} = b_0 + b_1
v^r_t,
\]
where $v^r_t =  \{ \frac{1}{2S} \sum_{s=1}^S\sum_{l=0}^{1}(y^r_{s,t-l} - y^r_{s,t-l-1})^2  \}^{1/2}$ and $b_0$, $b_1
> 0$.
\end{longlist}

The coefficients in the center parameter and scale parameter
models are estimated numerically by minimizing the continuous ranked
probability score (CRPS) for a truncated normal distribution, based on
a 45-day-sliding window; see \citet{gneiting2006} and \citet{gneiting2007}.

Two models are introduced briefly here as references:

\begin{longlist}[(ii)]
\item[(i)] PSS assumes the future wind speed is the same as the current
wind speed, $\hat{y}_{s,t+k} = y_{s,t}$.
\item[(ii)] As mentioned above, in the RSTD model [\citet{gneiting2006}], forecasting
regimes are defined based on the prevailing wind direction, and for each
regime a separate model is fitted only with historical wind speeds
as predictors in equation (\ref{geowindeq.mur}) plus speeds from
neighboring stations.
\end{longlist}

\subsection{The TDDGW model}

Based on the discussion of the geostrophic wind in Section~\ref{geowindsec.geowind}, it is clear that atmospheric pressure and
temperature play important roles in wind speed and direction. To reduce
the uncertainties in wind, an efficient short-term forecasting model
should include this critical information. However, the experiments in
the next section show that incorporating air pressure and temperature
directly into the TDD model does not reduce errors in forecasts. This
is because in the TDD model, particularly in the mean structure in
equation (\ref{geowindeq.mur}), linearity is assumed between future
wind speeds and the covariates. This assumption is invalid between wind
speed and air pressure or temperature. As a result, no improvement is
achieved by incorporating these variables directly into the TDD model.

Instead of seeking nonlinear forms between wind speeds and air pressure
and temperature in the mean structure of the TDD model, it is proposed
to use the geostrophic wind as a predictor, as this better expresses
the physical relationship between temperature, pressure and wind. In
the TDDGW model, geostrophic wind is incorporated into the TDD model,
hence keeping the model structure almost the same. Specifically, the
TDD model is modified by adding geostrophic wind into the center
parameter model in equation~(\ref{geowindeq.mur}):
%
\begin{eqnarray}
\label{geowindeq.murgw}
\mu^r_{1,t+k} &=& \alpha_0 + \sum
_{s=1,\ldots,S} \biggl[ \sum_{j=0,1,\ldots
,q_s}\alpha_{s,j} y^r_{s,t-j}\nonumber\\
&&\hspace{61pt}{}+ \sum_{j'=0,1,\ldots,q_s'} \bigl\{ \beta _{s,j'}\cos\bigl(
\theta^r_{s,t-j'}\bigr) + \gamma_{s,j'}\sin\bigl(\theta
^r_{s,t-j'}\bigr) \bigr\} \biggr]
\\
&&{}+ c_0 (w_g)^r_{1,t} +
c_1 (w_g)^r_{1,t-1}+ c_2
(w_g)^r_{1,t-2}+\cdots+ c_q
(w_g)^r_{1,t-q},\nonumber
\end{eqnarray}
where $q$ is the time lag of geostrophic wind depending on the station,
$s$, determined by the aforementioned modified BIC method and, again,
$w_g$ indicates the geostrophic wind speed. Since geostrophic wind is
the theoretical wind above the planetary boundary layer in the
atmosphere, its value for a small area is almost constant. This is why
the geostrophic wind is used as a common predictor in equation~(\ref{geowindeq.murgw}).

In addition to including geostrophic wind in the TDD model, the
geo\-strophic wind direction and the temperature difference between the
current and previous day are also considered, because, from the
atmospheric science point of view, these variables are closely related
to surface wind. These two modified TDDGW models are named TDDGWD and
TDDGWT, and with the two variables simultaneously, TDDGWDT.

Additionally, the diurnal pattern fitting is also modified. Instead of
the daily wind pattern in the model in equation (\ref{geowindeq.D}),
the average wind speed of each hour within a certain period is treated
as the diurnal pattern. Depending on the period used, there are several
versions of the diurnal pattern modeling: MD, a diurnal pattern that
takes into account winds in a 45-day-sliding window; SMD, a diurnal
pattern that is calculated for each season; and YMD, a diurnal pattern
based on a whole year's data (or several years' data).

\section{West Texas data}\label{geowindsec.data}

\subsection{Data description}

The wind data considered here were collected from mesonet towers at a
height of 10 m above the surface in West Texas and Eastern New Mexico,
and was also used by \citet{xie2014}. The original data archive contains
five-minute means of three-second measurements of wind and other
atmospheric parameters from more than 60 stations. In the experiment,
hourly-averaged data of five-minute means from 1 January 2008 to 31 December
2010 are used, divided into training data (2008--2009) and testing data
(2010). Although most wind turbine towers today are at least 60 m tall
[\citet{busby2012}], winds at
10 m height provide some information about the wind at turbine height
depending on the state of the planetary boundary layer. Moreover, we
are using 10 m winds because it is all that is available from this data set.

\begin{figure}

\includegraphics{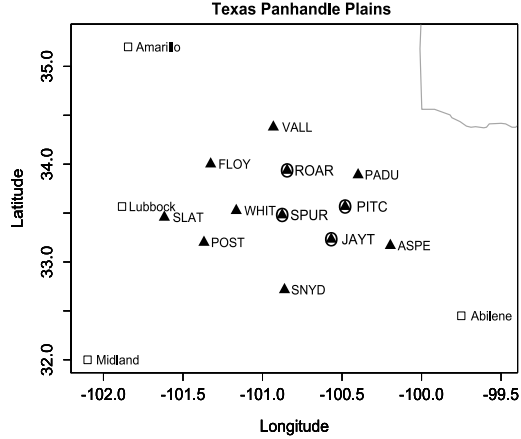}

\caption{The distribution of selected mesonet towers (triangles) in
West Texas (Panhandle plains). The four towers of PICT, JAYT, SPUR and
ROAR are marked by circled triangles. The 12 stations selected to
estimate the geostrophic wind are marked by triangles.}
\label{fig.datamap}
\end{figure}

\begin{figure}[b]

\includegraphics{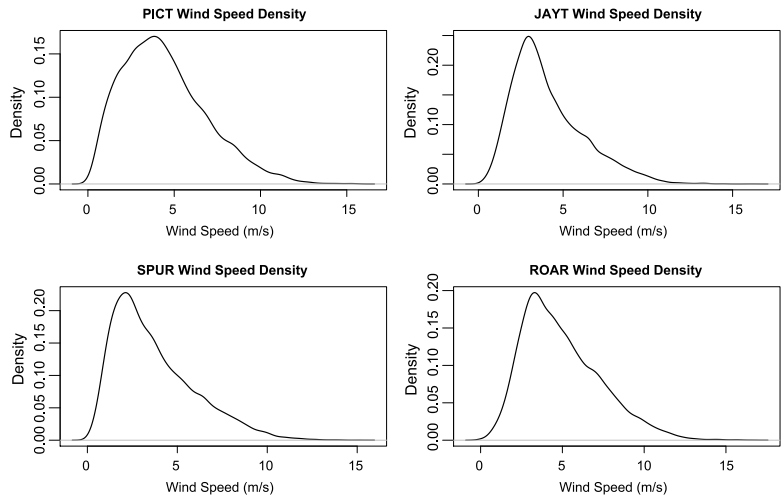}

\caption{Marginal density plots of wind speeds at PICT, JAYT, SPUR and
ROAR in 2008--2009.}
\label{fig.density}
\end{figure}

In our numerical experiment, a small area in the Panhandle plains is
chosen with four stations to test the newly proposed model; see Figure~\ref{fig.datamap}. This area includes PICT, JAYT, SPUR and ROAR
stations in and around Dickens county, between 40 to 55 miles apart
from one another. These four locations are marked by a circled triangle
in Figure~\ref{fig.datamap}. Our goal is to predict 1- to 6-hour-ahead
wind speeds at these four locations. The recorded data include wind
speed, wind direction, temperature and pressure. To estimate the
geostrophic wind in the TDDGW model, 12 surface stations were selected
(triangles in Figure~\ref{fig.datamap}) that surround the four test
stations. More information is given at \url{http://www.mesonet.ttu.edu/wind.html}.

The area where the four target stations are located in West Texas has
both northerly and southerly prevailing winds as shown by \citet{xie2014} with wind roses based on the 2008--2009 training data set.
High frequencies and large speed ranges are found from the north and
south directions at all four stations. More specifically, the southerly
wind dominates this area, with more frequent wind blowing from the
south than from the north. Different from the other three locations,
the station SPUR has a high frequency from the northwest direction. The
wind speed marginal density plots at the four stations are displayed in
Figure~\ref{fig.density} based on the wind data from 2008 and 2009.
They are positive and skewed to the right.

\subsection{Geostrophic wind and surface wind}

To estimate the geostrophic wind based on surface measurements of air
pressure and temperature, the aforementioned two steps in Section~\ref{geowindsec.geowind} are carried out. First, for each hour, surface
pressure measurements are represented by the geopotential height with
equation (\ref{hypsometric}). For the value of $\bar{T}$, the average
temperature from the 12 stations in Figure~\ref{fig.datamap} is
used and 850 hPa for the reference pressure, $p_{\mathrm{ref}}$. Second, using
the 12 stations' geopotential height data, along with their latitude
and longitude data, a geopotential height plane (\ref{plane}) is fitted
for each hour, resulting in a geopotential height gradient based on the
coefficients of the plane of the $x$ and $y$ horizontal components as
shown in equation (\ref{coefficients}). The monthly average
geopotential height is removed before fitting the plane. With these two
steps, each hourly surface wind record has a corresponding geostrophic
wind estimated from the temperature and pressure information.

\begin{figure}[b]

\includegraphics{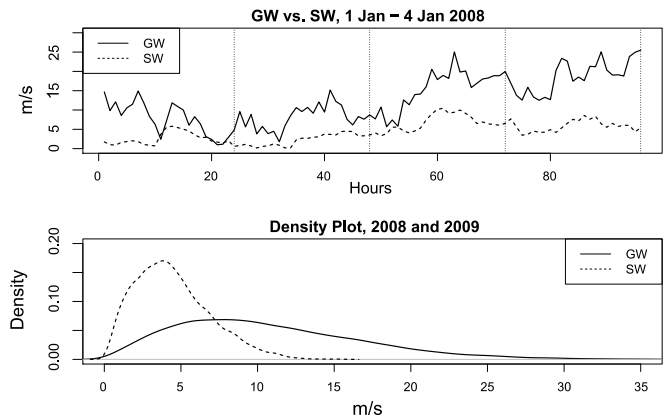}

\caption{Geostrophic wind (GW) vs. surface wind (SW) (top) and density
plots of the geostrophic wind and surface wind (bottom).}
\label{fig.gwplots}
\end{figure}
The four days' hourly geostrophic wind speeds (solid curve) and surface
winds (dashed curve) in 2008 at PICT in Figure~\ref{fig.gwplots} (top)
indicate that the former has larger values than the latter, while the
latter has larger amplitude of variation. Since the effects of friction
forces, which slow down the wind speed and change direction, are
ignored in the geostrophic balance, the geostrophic winds are stronger
and smoother than the surface winds. Also, it can be seen in Figure~\ref{fig.gwplots} that they share similar patterns, which is consistent
with the large positive correlation coefficient between the surface
wind and the geostrophic wind as listed in Table~\ref{geowindtab.cor}
in the next section. The bottom plot displays the density estimations
of the geostrophic wind speed (solid curve) and the surface wind speed
(dashed curve), from which we can see again that the geostrophic wind speed has a
larger range than does the surface wind speed.

\begin{figure}[t]

\includegraphics{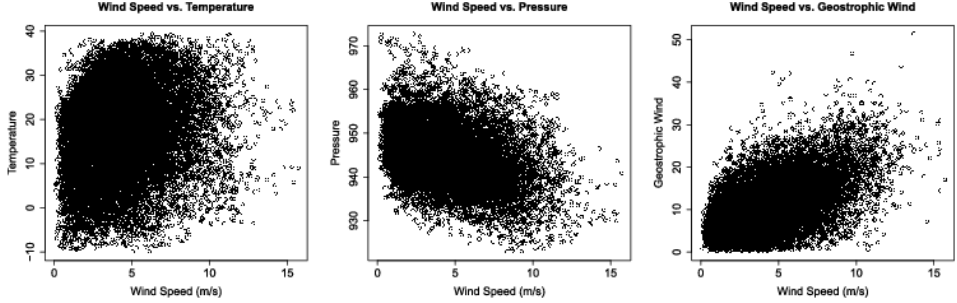}

\caption{Scatter plots of wind speed vs. temperature (Celsius) (left),
pressure (hPa) (middle) and geostrophic wind speed (m$/$s) (right).}
\label{fig.scatter}
\end{figure}

Figure~\ref{fig.scatter} displays scatter plots of wind speed vs.
surface temperature (left), pressure (middle) and geostrophic wind
speed (right) based on the training data at PICT. From the first plot,
we can see that the surface wind speed is very weakly correlated with
temperature. The correlation coefficient between them is $0.19$. The
correlation coefficient of the surface wind speed and pressure is
$-$0.34, indicating a weakly negative linear trend in the scatter plot
as well. However, the linearity correlation between surface wind and
geostrophic wind is stronger, with correlation coefficient equal to
$0.53$. This shows that geostrophic wind not only contains important
temperature and pressure information, but also meets the linearity
assumption such that it can be integrated into the TDD model. More
importantly, geostrophic wind has physical interpretability.
%

\begin{figure}[b]

\includegraphics{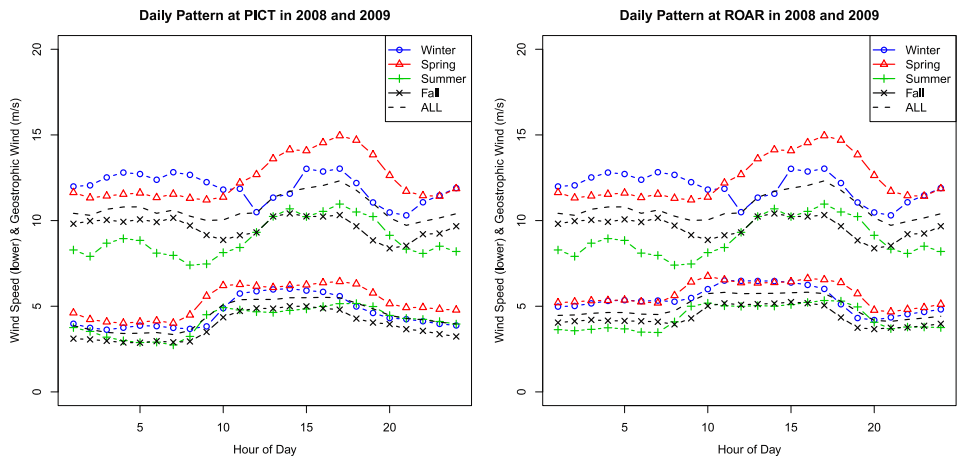}

\caption{Daily pattern of wind speed (lower part in each plot) and
geostrophic wind speed (upper part in each plot) in different seasons
of 2008--2009 at PICT (left) and ROAR (right).}
\label{fig.gwspd}
\end{figure}

Figure~\ref{fig.gwspd} shows the averaged diurnal pattern of the
surface wind speed and geostrophic wind in different seasons of
2008--2009 at PICT (left) and ROAR (right). The plots show that
geostrophic wind has higher speed than surface wind, which is slowed
down by the ground friction. The geostrophic wind for the two stations
is the same, but the surface winds are different albeit similar.
Through the hours of the day, the geostrophic wind fluctuates with a
range from 7 to 15~m$/$s, while the surface wind is smoother with a range
from 3 to 6 m$/$s. Seasonally, geostrophic wind and surface wind are
consistent, having higher speed during winter (December to February)
and spring (March to May) than summer (June to August) and fall
(September to November).

\section{Numerical results}\label{geowindsec.results}

\subsection{Training results}

\begin{table}[b]
\caption{Correlation coefficients between $y_{P,t+2}$ and the current
and up to 5-step lag surface wind speed~($y$), direction ($\theta$),
geostrophic wind speed ($w_g$) and geostrophic wind direction ($\theta
_g$) at~four~stations ($P$, $J$, $S$ and $R$)}
\label{geowindtab.cor}
\begin{tabular*}{\tablewidth}{@{\extracolsep{\fill}}@{}ld{2.2}d{2.2}d{2.2}d{2.2}d{2.2}d{2.2}@{}}
\hline
\textbf{Variable} & \multicolumn{1}{c}{$\bolds{t}$} & \multicolumn{1}{c}{$\bolds{t-1}$} & \multicolumn{1}{c}{$\bolds{t-2}$} & \multicolumn{1}{c}{$\bolds{t-3}$}&
\multicolumn{1}{c}{$\bolds{t-4}$} &
\multicolumn{1}{c@{}}{$\bolds{t-5}$}\\
\hline
$y_P$ & 0.80 & 0.70 & 0.62 & 0.54 & 0.47 & 0.40 \\
$w_{g,P}$ & 0.57 & 0.55 & 0.53 & 0.50 & 0.47 & 0.43 \\
$\cos(\theta_P)$& -0.06 & -0.08 & -0.11 & -0.13 & -0.15 &
-0.17 \\
$\sin(\theta_P)$&-0.14 & -0.16 & -0.17 & -0.19 & -0.20 &
-0.20 \\
$\cos(\theta_{g,P})$& 0.10 & 0.09 & 0.09 & 0.09 & 0.10 & 0.10 \\
$\sin(\theta_{g,P})$&0.17 & 0.15 & 0.13 & 0.11 & 0.09 & 0.07 \\[3pt]
$y_J$& 0.74 & 0.66 & 0.58 & 0.51 & 0.45 & 0.39 \\
$\cos(\theta_J)$&-0.12 & -0.14 & -0.15 & -0.16 & -0.17 &
-0.18 \\
$\sin(\theta_J)$&-0.14 & -0.17 & -0.19 & -0.21 & -0.22 &
-0.23 \\[3pt]
$y_S$ &0.73 & 0.64 & 0.55 & 0.48 & 0.40 & 0.33 \\
$\cos(\theta_S)$&-0.19 & -0.20 & -0.20 & -0.20 & -0.20 &
-0.20 \\
$\sin(\theta_S)$&-0.06 & -0.09 & -0.11 & -0.14 & -0.16 &
-0.18 \\[3pt]
$y_R$&0.76 & 0.70 & 0.64 & 0.59 & 0.53 & 0.48 \\
$\cos(\theta_R)$&-0.03 & -0.04 & -0.06 & -0.08 & -0.10 &
-0.12 \\
$\sin(\theta_R)$&-0.05 & -0.08 & -0.11 & -0.13 & -0.16 &
-0.17 \\
\hline
\end{tabular*}
\end{table}

In the training procedure the models for the center parameter are
obtained based on the training data set to forecast 1- to 6-hour-ahead
wind speed at each of the four stations. For example, to predict
$y_{P,t+2}$, the 2-hour-ahead wind speed at PICT, the variables listed
in Table~\ref{geowindtab.cor}, except geostrophic wind direction, are
put into the selection pool, and the aforementioned BIC is applied to
select significant predictors. The variables in the selection pool
include current and up to 10-step lags of wind speed, geostrophic wind
speed, and pairs of cosine and sine of the wind direction at all four
stations. In the TDDGWD model, the cosine and sine of the geostrophic
wind direction are also considered. Different from the cosine and sine
of the surface wind direction, which have negative correlations with
the 2-hour-ahead wind speed at PICT, the cosine and sine of the
geostrophic wind direction are positively correlated with the 2-hour-ahead
wind speed at PICT (see Table~\ref{geowindtab.cor}). In the table, the
indexes, $P$, $J$, $S$ and $R$, indicate the four locations.

\subsection{Evaluation of forecasts}

The trained TDDGW, TDDGWT,\break TDDGWD and TDDGWDT models are applied to the
testing data set with modified diurnal modeling, MD, SMD and YMD, to
predict probabilistically 1- to 6-hour-ahead wind speeds at the four
stations. Prediction mean absolute errors (MAE) are used to evaluate
the performance of the forecasts, which are defined as $\sum_{t=1}^T|y_{P,t+2}-\hat{y}_{P,t+2}|$, at station PICT for 2-hour-ahead,
for example. When $\hat{y}_{P,t+2}$ equals to the median of the
predictive distribution, the error reaches the minimum value. Thus, for
the truncated normal distribution, we take the median as forecast:
\[
\hat{y}_{P,t+2} = \mu_{P,t+2} + \sigma_{P,t+2} \cdot
\Phi^{-1}\bigl\{0.5+0.5 \cdot\Phi(-\mu_{P,t+2}/
\sigma_{P,t+2})\bigr\};
\]
see \citet{gneiting2011} for a discussion of quantiles as optimal point
forecasts. A~45-day-sliding window is used to estimate the coefficients
in the models with the CRPS method. Forecasts are compared with the
reference models listed in Section~\ref{refforc} in addition to the
TDD model.

Besides MAE, the RMSE and CRPS are also used to compare model
performance. Compared with MAE, RMSE has stronger penalty on large
forecast errors. CRPS essentially provides a measure of probabilistic
forecast performance. The computation of the CRPS for the truncated
normal distribution can be found in \citet{gneiting2006}.

%
\begin{table}[b]
\tabcolsep=0pt
\caption{MAE values (m$/$s) of 2-hour-ahead forecasts from TDDGW with
different diurnal component fitting methods at PICT in 2010. The
smallest MAE value of each column is boldfaced}
\label{geowindtab.mdiurnal}
\begin{tabular*}{\textwidth}{@{\extracolsep{\fill}}@{}llccccccccccccc@{}}
\hline
&&&&&&&&&&&&&&\textbf{Ove-}
\\
\textbf{Site} & \textbf{Model} & \textbf{Jan.} & \textbf{Feb.} & \textbf{Mar.} & \textbf{Apr.} & \textbf{May} & \textbf{Jun.} & \textbf{Jul.} & \textbf{Aug.} & \textbf{Sep.} &
\textbf{Oct.} & \textbf{Nov.} & \textbf{Dec.} & \textbf{rall} \\
\hline
PICT & TDDGW & 0.95 & 0.81 & 1.02 & 0.93 & $\bolds{0.91}$ & 0.96 & 0.91 &
0.91 & 0.84 & 0.82 & 0.97 & 0.98 & 0.92 \\
PICT & TDDGW-MD & $\bolds{0.94}$ & $\bolds{0.80}$ & $\bolds{0.96}$ & 0.89 & 0.92 &
$\bolds{0.91}$ & $\bolds{0.83}$ &$\bolds{0.86}$ & 0.81 &$\bolds{0.77}$ & $\bolds{0.92}$
&$\bolds{0.94}$ &$\bolds{0.88}$ \\
PICT & TDDGW-SMD & $\bolds{0.94}$ & 0.84 & 0.98 & $\bolds{0.88}$ & 0.93 & 0.93
& 0.85 & $\bolds{0.86}$ & 0.82 & 0.78 & 0.94 & 0.96 & 0.89 \\
PICT & TDDGW-YMD & 0.98 & 0.81 & 0.98 & 0.91 & 0.95 & 0.96 & 0.86 &
0.88 & $\bolds{0.80}$ & 0.81 & 0.96 & 0.98 & 0.90 \\
\hline
\end{tabular*}
\end{table}

In Table~\ref{geowindtab.mdiurnal} the prediction MAE values of
2-hour-ahead forecasts at PICT in 2010 from the TDDGW model with
aforementioned different diurnal modeling methods are listed. Overall,
the MD method has the smallest MAE values among the four, 0.88 m$/$s
compared with 0.92 m$/$s, 0.89 m$/$s and 0.90 m$/$s, from TDDGW, TDDGW-SMD
and TDDGW-YMD methods, respectively. The TDDGW-MD model has the
smallest MAE values, 10 out of the 12 months, followed by TDDGW-SMD, 3
out of 12 months.

The modified methods fit the diurnal pattern better than the one in
equation (\ref{geowindeq.D}). This is because the latter fits the
pattern by a continuous smooth function of the time of a day. The
fitted results would be adjusted to the average wind speed of the day,
while MD, SMD and YMD only provide the average wind speed on the hours.
Since the focus is on hourly ahead forecasting, here using MD, SMD and
YMD is reasonable without losing functionality in practice. Therefore,
in the following only forecasts from models that use the MD method to
fit the diurnal component are displayed.

\begin{table}[b]
\tabcolsep=1.2pt
\caption{MAE, RMSE and CRPS values (m$/$s) of 2-hour-ahead forecasts from
various forecasting models at PICT in 2010. The smallest value of each
criteria in each column is boldfaced}
\label{geowindtab.maeres}
\begin{tabular}{@{}llccccccccccccc@{}}
\hline
& & & & & & & & & & & & & & \textbf{Ove-}
\\
\textbf{Site} & \textbf{Model} & \textbf{Jan.} & \textbf{Feb.} & \textbf{Mar.} & \textbf{Apr.} & \textbf{May} & \textbf{Jun.} & \textbf{Jul.} & \textbf{Aug.} & \textbf{Sep.} &
\textbf{Oct.} & \textbf{Nov.} & \textbf{Dec.} & \textbf{rall} \\
\hline
MAE& &&&&&&&&&&&&\\
PICT & PSS & 1.06 & 0.87 & 1.21 & 1.15 & 1.15 & 1.13 & 1.03 & 1.05 &
0.96 & 0.97 & 1.17 & 1.14 & 1.08 \\
PICT & RSTD & 0.93 &$\bolds{0.79}$ & 1.07 & 0.98 & 1.02 & 0.95 & 0.89 &
0.90 & 0.83 & 0.82 & 0.99 & 1.01 & 0.94 \\
PICT & TDD & 0.95 & 0.81 & 1.07 & 0.99 & 1.00 & 0.97 & 0.89 & 0.93 &
0.84 & 0.86 & 1.01 & 1.03 & 0.95 \\
PICT & TDDGW-MD & 0.94 & 0.80 & $\bolds{0.96}$ &$\bolds{0.89}$ &$\bolds{0.92}$ &
0.91 &$\bolds{0.83}$ &$\bolds{0.86}$ & 0.81 &$\bolds{0.77}$ &$\bolds{0.92}$ & $\bolds{0.94}$ &$\bolds{0.88}$ \\
PICT & TDDGWT-MD & 0.94 & 0.82 &$\bolds{0.96}$ & 0.90 &$\bolds{0.92}$ & 0.91
&$\bolds{0.83}$ &$\bolds{0.86}$ & 0.81 &$\bolds{0.77}$ &$\bolds{0.92}$ & 0.95 &$\bolds{0.88}$ \\
PICT & TDDGWD-MD &$\bolds{0.91}$ & 0.82 & 0.97 &$\bolds{0.89}$ &$\bolds{0.92}$
&$\bolds{0.90}$ & 0.84 &$\bolds{0.86}$ &$\bolds{0.80}$ & 0.78 &$\bolds{0.92}$ & $\bolds{0.94}$ &$\bolds{0.88}$ \\
PICT & TDDGWDT-MD &$\bolds{0.91}$ & 0.83 & 0.97 & 0.90 &$\bolds{0.92}$ &$\bolds{0.90}$ & 0.84 &$\bolds{0.86}$ &$\bolds{0.80}$ & 0.78 &$\bolds{0.92}$ &$\bolds{0.94}$
&$\bolds{0.88}$ \\[3pt]
RMSE& &&&&&&&&&&&&\\
PICT & PSS & 1.45 & 1.19 & 1.66 & 1.54 & 1.52 & 1.52 & 1.40 & 1.44 &
1.30 & 1.31 & 1.63 & 1.56 & 1.47 \\
PICT & RSTD & 1.25 & $\bolds{1.05}$ & 1.40 & 1.29 & 1.35 & 1.25 & 1.20 &
1.16 & 1.09 & 1.07 & 1.30 & 1.35 & 1.24 \\
PICT & TDD & 1.25 & 1.07 & 1.41 & 1.28 & 1.31 & 1.26 & 1.19 & 1.21 &
1.09 & 1.09 & 1.31 & 1.38 & 1.25 \\
PICT & TDDGW-MD & 1.23 & $\bolds{1.05}$ & $\bolds{1.28}$ & $\bolds{1.15}$ & $\bolds{1.22}$
& 1.22 & 1.13 & $\bolds{1.12}$ & 1.05 & $\bolds{1.02}$ & 1.21 & $\bolds{1.28}$ &
$\bolds{1.17 }$\\
PICT & TDDGWT-MD & 1.23 & 1.07 & $\bolds{1.28}$ & $\bolds{1.15}$ &$\bolds{1.22}$ & $\bolds{1.21}$ & $\bolds{1.12}$ & $\bolds{1.12}$ & 1.05 & $\bolds{1.02}$ &
$\bolds{1.20}$ & $\bolds{1.28}$ & $\bolds{1.17 }$\\
PICT & TDDGWD-MD & $\bolds{1.20}$ & 1.08 & 1.29 & $\bolds{1.15}$ & 1.23 & 1.22 & $\bolds{1.12}$ & 1.13 & $\bolds{1.03}$ & $\bolds{1.02}$ & 1.21 & $\bolds{1.28}$ &
$\bolds{1.17}$\\
PICT & TDDGWDT-MD & 1.21 & 1.09 & 1.29 & 1.16 & 1.23 & $\bolds{1.21}$ & $\bolds{1.12}$ & 1.13 & $\bolds{1.03}$ & $\bolds{1.02}$ & 1.21 & $\bolds{1.28}$ &
$\bolds{1.17}$\\[3pt]
CRPS& &&&&&&&&&&&&\\
PICT & RSTD & 0.67 & $\bolds{0.57}$ & 0.77 & 0.71 & 0.73 & 0.68 & 0.65 &
0.64 & 0.59 & 0.59 & 0.71 & 0.72 & 0.67 \\
PICT & TDD & 0.68 & 0.58 & 0.77 & 0.71 & 0.71 & 0.69 & 0.64 & 0.67 &
0.60 & 0.61 & 0.72 & 0.74 & 0.68 \\
PICT & TDDGW-MD & 0.66 & 0.58 & $\bolds{0.70}$ & $\bolds{0.64}$ & $\bolds{0.66}$
& $\bolds{0.65}$ & $\bolds{0.60}$ & 0.62 & 0.58 & 0.56 & $\bolds{0.66}$ & $\bolds{0.68}$ & $\bolds{0.63 }$\\
PICT & TDDGWT-MD & 0.66 & 0.59 & $\bolds{0.70}$ & $\bolds{0.64}$ & $\bolds{0.66}$ & $\bolds{0.65}$ & $\bolds{0.60}$ & $\bolds{0.61}$ & 0.58 & $\bolds{0.55}$ &
$\bolds{0.66}$ & $\bolds{0.68}$ & $\bolds{0.63 }$\\
PICT & TDDGWD-MD & $\bolds{0.65}$ & 0.59 & $\bolds{0.70}$ & $\bolds{0.64}$ &
$\bolds{0.66}$ & $\bolds{0.65}$ & $\bolds{0.60}$ & 0.62 & $\bolds{0.56}$ & 0.56 &
0.67 & $\bolds{0.68}$ & $\bolds{0.63 }$\\
PICT & TDDGWDT-MD & $\bolds{0.65}$ & 0.60 & $\bolds{0.70}$ & $\bolds{0.64}$ & $\bolds{0.66}$ & $\bolds{0.65}$ & $\bolds{0.60}$ & 0.62 & 0.57 & 0.56 & 0.67 &
$\bolds{0.68}$ & $\bolds{0.63 }$\\
\hline
\end{tabular}
\end{table}

The MAE, RMSE and CRPS values of 2-hour-ahead forecasts from
different models at PICT in 2010 are listed in Table~\ref{geowindtab.maeres}. At PICT, it can be observed that all the
space--time models outperform the PSS model as expected, with smaller
MAE values. Except for February, our new models that incorporate
geostrophic wind give more accurate forecasts than the RSTD and TDD
models do, with the MAE value 0.88 m$/$s compared with 0.94 m$/$s and 0.95
m$/$s. Up to
two decimal points, the TDDGW-MD, TDDGWT-MD, TDDGWD-MD and TDDGWDT-MD
models have similar MAE values, around 0.88 m$/$s. Looking more closely, the
TDDGWD-MD gives the largest reduction in the relative MAE value,
around $18.3\%$. As expected, the models including geostrophic wind
are better than the other two
space--time models (RSTD and TDD) with 13.2\% and 12.1\% reductions in
MAE values relative to PSS. Comparing the results of the TDDGW and
TDDGW-MD models, the modified diurnal pattern modeling based on the
45-day-sliding window helps to provide a 3.7\% reduction in the MAE
value relative to PSS. Similar results can be seen based on CRPS and
RMSE. Our new models produce the smallest CRPS and RMSE.

Looking across the 4 locations, our new method obtains between 13.9\%
and 22.4\% overall mean absolute error reduction relative to
persistence in 2-hour-ahead forecasts,
and between 5.3\% and 8.2\% reduction relative
to the best previous space--time methods in this setting.

\begin{figure}

\includegraphics{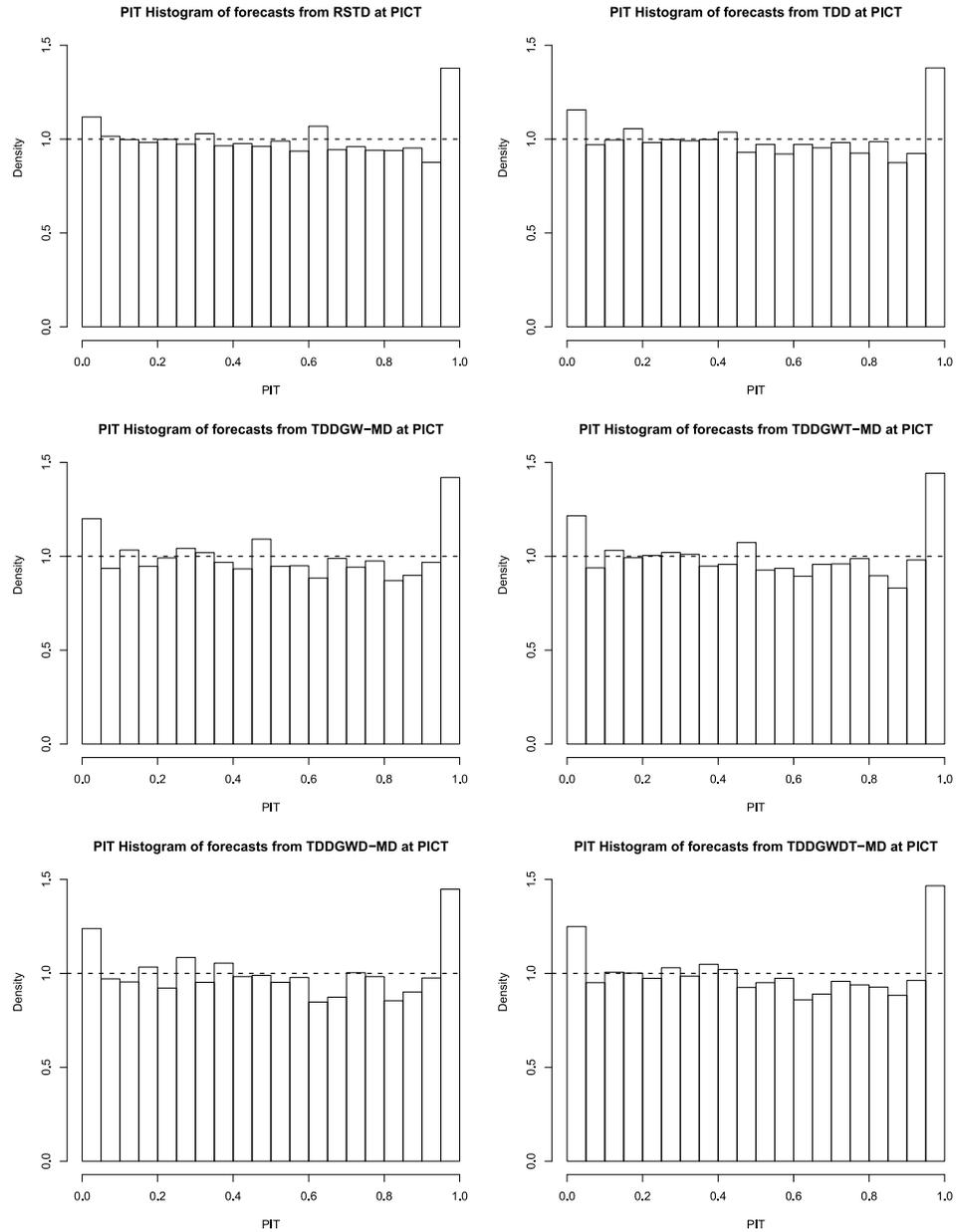}

\caption{PIT histograms for RSTD, TDD, TDDGW-MD, TDDGWT-MD, TDDGWD-MD
and \mbox{TDDGWDT}-MD predictive distributions of 2-hour-ahead forecasts at
PICT in\vspace{-2pt} 2010.}
\label{fig.PIT}
\end{figure}

To assess calibration, we display the histograms of the probability
integral transform (PIT)
of our models in Figure~\ref{fig.PIT} for 2-hour-ahead forecasts at
PICT in 2010. The PIT is the value attained by the predictive distribution
at the observation [\citet{daw1984,dieb1998}]. We see that all six PIT
histograms are approximately
uniform, hence indicating calibration and prediction intervals that
have close to nominal coverage at all levels.
To assess sharpness, we compute the average width of the 90\% central
prediction intervals.
We obtain 3.59 m$/$s for RSTD and 3.61 m$/$s for TDD, whereas for the
models based on geostrophic wind
the values drop to between 3.34 m$/$s for TDDGW-MD and 3.29~m$/$s for
TDDGWDT-MD, a reduction of forecast uncertainty of about
\mbox{7--9\%}.

%
\begin{figure}

\includegraphics{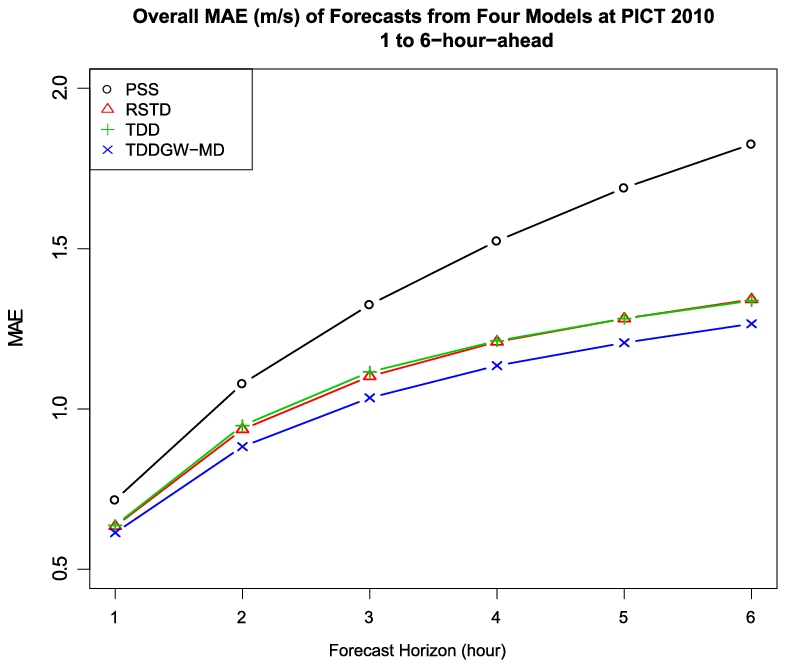}

\caption{Plot of MAE (m$/$s) of forecasts from the PSS, RSTD, TDD   and
TDDGW-MD models for 1- to 6-hour-ahead forecasting at PICT in 2010.}
\label{fig.1to6}
\end{figure}

Figure~\ref{fig.1to6} displays overall MAE values of forecasts from
the PSS, RSTD, TDD  and TDDGW-MD models for 1- to
6-hour-ahead forecasting at PICT in 2010. We can see that the
space--time models improve the forecasting accuracy with smaller MAE
values compared to PSS. Our new model has smallest MAE values for all
the six forecast horizons. The RSTD and TDD   have quite close results. At
the same time, as expected, forecasting accuracy decreases with the
increase of forecasting horizon for all the four models, but the
space--time models have a~smaller increasing rate than the PSS model.
Similar results were obtained with RMSE and CRPS, as well as at other locations.

\section{Final remarks} \label{geowindsec.conclusion}

Accurate wind prediction is critical in running power systems that
have large shares of wind power. In recent decades many studies have
been devoted to improving short-term wind forecasting for large-scale
wind power development around the world.

This paper developed statistical short-term wind forecasting models
based on atmospheric dynamics principles. It proposed the use of the
geostrophic wind as a predictor. The geostrophic wind is a good
approximation to the winds in the extratropical free troposphere and
can be estimated using only surface pressure and temperature data. In terms
of the underlying atmospheric physics, the geostrophic wind is
correlated to the real wind more strongly than either temperature or
pressure. This is demonstrated by the fact that no improvement was
found by directly incorporating atmospheric temperature and pressure
into the most advanced space--time forecasting model to date. The
geostrophic wind can be approximated from networks of standard surface
meteorological observations. More importantly, it helps to reduce
prediction errors significantly when incorporated into space--time
models.

In this paper, more accurate forecasts were achieved by incorporating
geo\-strophic wind as a predictor into space--time statistical models and
modifying diurnal pattern models in 1- to 6-hour-ahead wind speed
forecasting. In addition, trigonometric functions of the geostrophic
wind direction
and temperature differences between the current and previous day were
also considered. We showed how simpler but more efficient methods can
be applied to
fit the diurnal pattern of wind to obtain better forecasts.

With our new model and existing space--time models, we forecast the
1- to 6-hour-ahead wind speeds at four locations in West
Texas. Three different criteria were used to evaluate the performance
of the models, including MAE, RMSE and CRPS. The results showed that
our new models outperform the PSS, RSTD and TDD   in
terms of all three criteria. Moreover, PIT histograms confirmed that
our new models based
on geostrophic wind were calibrated and sharp.
\citet{xie2014} further quantified
the overall cost benefits on power system dispatch by
reducing uncertainties in near-term wind speed forecasts
based on the TDDGW model.

\section*{Acknowledgments}
The authors thank the West Texas Mesonet
(\surl{www.\\mesonet.ttu.edu}), operated by Texas Tech University, for use of
the mesonet meteorological data.

%



\printaddresses
\end{document}